\documentclass[twocolumn]{aastex701}
\usepackage{multirow}
\usepackage{amsmath}
\begin{document}

\title{Pulsar Gleaners\protect\footnote{The term ``gleaners'' in the title ``Pulsar Gleaners'' originates from the 1857 oil painting \emph{The Gleaners} by Jean-Fran\c{c}ois Millet, which depicts three peasant women collecting stray stalks of wheat left in a field after the harvest.}: Discovery of 19 Pulsars in FAST Archival Data at $|b|<5\degr$ and Decl.$<-5\degr$}

\shorttitle{Discovery of 19 Pulsars in FAST Archival Data}
\shortauthors{S.-J. Gao, Y.-X. Shao \& X.-D. Li}
\correspondingauthor{Shi-Jie Gao \& Xiang-Dong Li}

\author[orcid=0000-0002-0822-0337,sname='Gao']{Shi-Jie Gao}
\affiliation{School of Astronomy and Space Science, Nanjing University, Nanjing, 210023, People's Republic of China}
\affiliation{Key Laboratory of Modern Astronomy and Astrophysics, Nanjing University, Ministry of Education, Nanjing, 210023, People's Republic of China}
\email[show]{gaosj@nju.edu.cn}  

\author[orcid=0000-0001-5684-0103,sname='Shao']{Yi-Xuan Shao}
\affiliation{School of Astronomy and Space Science, Nanjing University, Nanjing, 210023, People's Republic of China}
\affiliation{Key Laboratory of Modern Astronomy and Astrophysics, Nanjing University, Ministry of Education, Nanjing, 210023, People's Republic of China}
\email[hide]{yixuan@smail.nju.edu.cn}

\author[orcid=0000-0002-0584-8145,sname='Li']{Xiang-Dong Li}
\affiliation{School of Astronomy and Space Science, Nanjing University, Nanjing, 210023, People's Republic of China}
\affiliation{Key Laboratory of Modern Astronomy and Astrophysics, Nanjing University, Ministry of Education, Nanjing, 210023, People's Republic of China}
\email[show]{lixd@nju.edu.cn}

\begin{abstract}
We report the discovery of 19 new pulsars identified from archival observations of the Five-hundred-meter Aperture Spherical radio Telescope (FAST) within Galactic latitudes $|b|<5\degr$ and declinations ${\rm Decl.}<-5\degr$. The dataset was recorded using FAST's $L$-band 19-beam receiver and covered $\sim 3.6\deg^2$ with a cumulative integration time of $\sim$ 500 hr and a total raw data volume of $\sim$ 700 TB. Our search employed fast Fourier transform (FFT)-based and fast folding algorithm (FFA)-based periodic searches, and the single-pulse search. These new pulsars have spin periods range from 0.03 to 5.54~s. Two have periods under 0.1~s, suggesting they are likely young pulsars or mildly recycled pulsars. Four pulsars exhibit dispersion measures (DMs) exceeding $1000~{\rm pc~cm^{-3}}$ with PSR~J1839$-$0558t having the highest value in our sample at $\sim 1271~{\rm pc~cm^{-3}}$, providing valuable samples for pulsar studies in the high-DM regime. Two rotating radio transients, PSRs~J1836$-$0552t and J1847$-$0624t, were detected by FFA and single-pulse searches but failed with the FFT-based searches. In addition, three faint pulsars that were also missed by FFT-based searches were successfully detected using FFA. These discoveries demonstrate the critical role of FFA in uncovering faint, long-period, and sporadic pulsars, and highlight the significant potential of FAST archival data, especially when combined with longer integration times and complementary search techniques, to reveal rare and weak pulsar populations.
\end{abstract}


\section{Introduction}\label{sec:intro}

Pulsars are highly magnetized rotating neutron stars that emit beams of electromagnetic radiation (primarily radio waves) \citep{Hewish+1968}. They serve as powerful astrophysical laboratories, enabling studies of the neutron star physics \citep{hpa}, the interstellar medium and Galactic magnetic fields \citep[e.g.,][]{YMW16,Han+2018}, binary and stellar evolution \citep[see Chapter~14 of][]{pbse}, nano-Hz gravitation waves through pulsar timing arrays \citep[e.g.,][]{Agazie+2023,EPTA,Reardon+2023,Xu+2023}, and precision tests of general relativity \citep[e.g.,][]{Damour+1992,Kramer+2006}. To date more than four thousand pulsars have been discovered, as cataloged in the ATNF pulsar catalogue\footnote{Version 2.7.0, \url{https://www.atnf.csiro.au/research/pulsar/psrcat/}} \citep{ATNF}, yet several key pulsar populations remain poorly sampled or entirely unexplored, including sub-millisecond pulsars \citep{Burderi+2001,Du+2009}, double pulsar systems \citep{Burgay+2003}, pulsar-black hole binary systems \citep{Shao+2018,Chattopadhyay+2021,Barr+2024} and extremely long-period pulsars \citep[][]{Caleb+2022}. Continued pulsar searches are therefore essential for expanding the known population, uncovering rare systems, and opening new windows into fundamental physics.

The Five-hundred-meter Aperture Spherical radio Telescope \citep[FAST,][]{Nan+2006,Nan+2008,Nan+2011} is currently the most sensitive single-dish telescope in the world, enabling the detection of extremely faint pulsars. To date, FAST has discovered more than one thousand pulsars, as summarized in the News and Views article by \cite{Lorimer+2025}. It's major pulsar surveys includes the FAST Galactic Plane Pulsar Snapshot survey \citep[GPPS,][]{Han+2021,Han+2025}, the Commensal Radio Astronomy FAST Survey \citep[CRAFTS,][]{LiDi+2018} and the FAST Globular Cluster Pulsar Survey \citep[GC FANS,][]{Pan+2021,Lian+2025}. The GPPS survey covers the Galactic plane region with $|b|<10\degr$, utilizing the full gain of FAST at zenith angles $\lesssim 28.5\degr$ \citep{Han+2025} with an integration time of 5~min. However, many FAST accessible sky regions remain underexplored, particularly those at low Galactic latitudes and southern declinations. For Decl.$\lesssim -5\degr$, observations require backward illumination when the zenith angle exceeds $\sim 26\degr$ \citep{Jin+2013}. The snapshot mode adopted in the GPPS survey, which relies on rapidly repositioning the receivers within the cabin, is not applicable in this regime. This limitation arises mainly because the feed cabin needs to be rotated about its phase center toward the 500-m aperture to avoid seeing noise from the surroundings \citep{Jin+2013}. The CRAFTS survey is carried out in drift-scan mode, resulting in an effective integration time of only $\sim$10~s per pointing, and is therefore significantly less sensitive than the GPPS survey.

Normal observations at $\text{Decl.}<-5\degr$, including pulsar timing, polarization studies of pulsars, interplanetary scintillation observations and follow-up observations of compact objects, have been carried out extensively with FAST. The accumulated archival data spanning Data Releases 1--23\footnote{\url{https://fast.bao.ac.cn/cms/category/announcement}} therefore represent a valuable but largely untapped resource for new pulsar discoveries. Moreover, because FAST observations in this period were made with the 19-beam $L$-band receiver \citep{LiDi+2018,Jiang+2020}, the multibeam mode was often enabled during observations for radio-frequency interference (RFI) checks or for other operational reasons. This significantly increases the effective sky coverage of the archival data, making systematic pulsar searches feasible despite the observations not being originally designed for pulsar search purposes.

Two methods are commonly used to search for periodic pulsar signals: fast Fourier transform (FFT)-based techniques in the frequency domain, and the fast folding algorithm \citep[FFA,][]{Staelin+1969} in the time domain. FFT-based searches detect pulsars via incoherent harmonic summing in the frequency domain, requiring relatively low computational costs, and have led to many of the large surveys' pulsar discoveries. In contrast, the FFA is a phase-coherent time-domain method that folds dedispersed time-series directly, offering substantially greater sensitivity to periodic signals than the standard FFT incoherent harmonic summation, especially for long periods, nulling or narrow duty cycles \citep{Cameron+2017,Morello+2020,Singh+2023,Grover+2024}. In recent years, renewed interest in FFA has led to successful identification of exotic long-period pulsars in globular cluster M15 \citep{ZhouDK+2023,Wu+2023}, a millisecond pulsar in a binary system \citep{Li+2025}, and other pulsar discoveries \citep[e.g.,][]{Tyulbashev+2024}. Hence, incorporating an FFA-based search as a complement to FFT is now considered essential for comprehensive pulsar surveys. In addition to periodic search techniques, single-pulse searches provide a complementary approach for identifying single sporadic and bright radio pulses \citep{Cordes+2003}. This method has enabled the discovery of the first rotating radio transients \citep[RRATs,][]{McLaughlin+2006} and the first fast radio burst \citep{Lorimer+2007}, and remains essential for detecting transient or intermittently emitting pulsars that may be missed by periodic searches.

In this work, we present the discovery of 19 new pulsars identified from FAST archival observations at $|b|<5\degr$ and $\rm Decl.<-5\degr$. We employed both FFT-based and FFA-based periodic searches, as well as single-pulse search. The data selection, processing procedures, and search methodologies are described in Section~\ref{sec:obs}. The properties of the newly discovered pulsars, along with related discussions are presented in Section~\ref{sec:res}. We provide our conclusion and a summary in Section~\ref{sec:sum}.

\section{Methods}\label{sec:obs}

\begin{figure*}
    \centering
    \includegraphics[width=\linewidth]{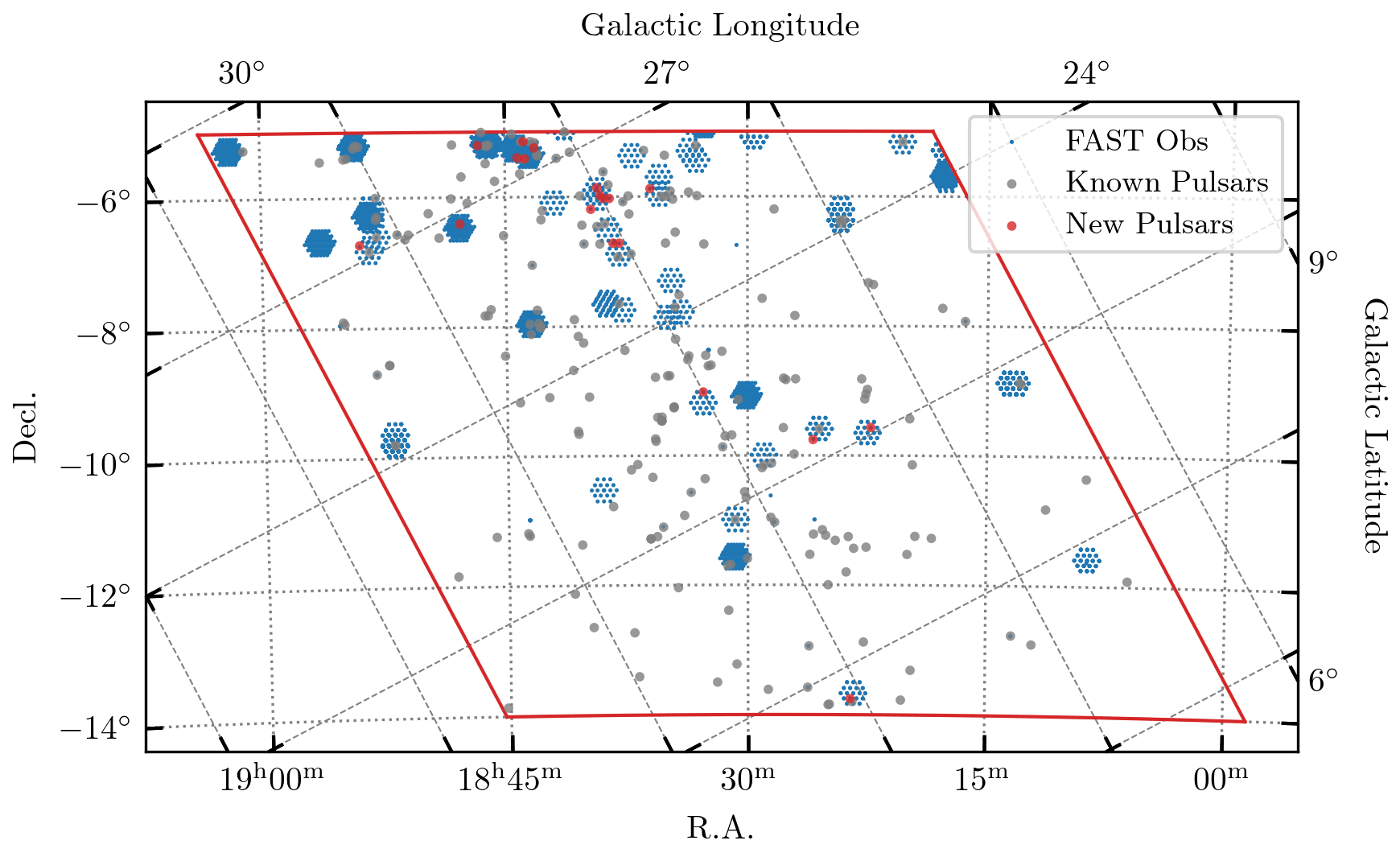}
    \caption{FAST observations used in this study. Data from FAST Data Releases 1--23 were selected, restricting the sample to observations with declination $\text{Decl.}<-5\degr$ and Galactic latitudes $|b|<5\degr$ (indicated by red solid lines). The $3'$ beam size is not to scale. Known pulsars are marked with gray dots, and new pulsars discovered in this work are marked with red dots.\label{fig:obs}}
\end{figure*}

\begin{table*}
    \centering
    \caption{Summary of the 19 newly discovered pulsars. For each source, we report the R.A. and Decl. of the beam center in which the pulsar was detected, the observing epoch (MJD), the integration time $T_{\rm obs}$, and the search method(s) that yielded a detection (Y for Yes and N for No): FFT-based periodic search (FFT), FFA-based periodic search (FFA), and single-pulse search (SP). We also list the measured (topocentric) spin period $P$, dispersion measure (DM), mean flux density, scatter-broadening timescale $\tau$, and distance estimates $d_{\rm DM}$ based on the YMW16 (former) and NE2001 (latter) electron density models. Owing to the $\sim3'$ beam size of the FAST 19-beam receiver, the true pulsar position is uncertain by approximately $\sim1.5'$ with respect to the beam center. To indicate this positional uncertainty, we append the suffix ``t'' (for ``temporary'') to their names. Pulsars confirmed with a second FAST observation are marked with an asterisk (*) before their names.\label{tab:psrs}}
    \begin{tabular}{ccccccccccccc}
\hline
PSR&R.A.&Decl.&MJD&$T_{\rm obs}$&FFT&FFA&SP&$P$&DM&Flux&$\tau$&$d_{\rm DM}$\\

&($\rm^h$:$\rm^m$:$\rm^s$)&($\degr$:$'$:$''$)&(day)&(min)&&&&(s)&($\rm pc~cm^{-3}$)&($\mu$Jy)&(ms)&(kpc)\\
\hline
*J1822$-$0933t &18:22:17.1 &$-$09:34:12 & 59924.24 &26.0 &Y &Y &N &0.3155 &349.2 &7.8 &38.5 &8.4, 5.7\\
*J1823$-$1345t &18:23:31.4 &$-$13:44:51 & 60299.21 &21.5 &Y &Y &N &0.4677 &1104.0 &17.0 &43.4 &6.1, 11.0\\
*J1825$-$0945t &18:25:53.7 &$-$09:45:02 & 60239.37 &46.0 &N &Y &N &2.3007 &533.4 &1.4 &13.3 &9.9, 6.9\\
*J1832$-$0901t &18:32:46.5 &$-$09:01:30 & 60594.39 &110.0 &Y &N &N &0.0297 &235.2 &2.1 &0.0 &3.7, 4.3\\
J1836$-$0552t &18:36:03.4 &$-$05:52:45 & 59264.02 &1.8 &N &Y &Y &5.5378 &336.2 &2.4 &4.0 &4.7, 5.6\\
J1837$-$0643t &18:37:58.0 &$-$06:43:12 & 60231.38 &90.0 &Y &Y &N &2.0491 &699.5 &11.4 &330.0 &5.3, 8.0\\
J1838$-$0601At &18:38:33.9 &$-$06:01:35 & 60247.34 &130.0 &Y &Y &N &1.0185 &1217.2 &4.8 &496.0 &6.7, 11.8\\
J1838$-$0601Bt &18:38:56.9 &$-$06:01:38 & 60247.34 &105.0 &Y &Y &N &1.9261 &1161.2 &5.4 &613.9 &6.5, 11.3\\
J1838$-$0643t &18:38:21.2 &$-$06:43:13 & 60231.38 &90.0 &Y &Y &N &3.3702 &561.6 &2.0 &3.0 &4.9, 7.0\\
J1839$-$0551t &18:39:20.0 &$-$05:51:55 & 60247.34 &130.0 &Y &Y &N &1.1742 &59.8 &1.1 &11.4 &1.7, 1.8\\
J1839$-$0558t &18:39:08.4 &$-$05:58:43 & 59632.06 &44.0 &Y &Y &N &1.9254 &1271.0 &6.4 &569.2 &6.7, 12.5\\
J1839$-$0611t &18:39:42.9 &$-$06:11:32 & 60247.34 &108.0 &Y &Y &N &4.9520 &868.0 &2.1 &4.9 &5.8, 9.3\\
*J1843$-$0508t &18:43:52.4 &$-$05:08:56 & 59848.51 &7.5 &Y &Y &N &0.2515 &152.6 &10.9 &0.5 &3.4, 3.3\\
J1843$-$0514t &18:43:12.6 &$-$05:14:43 & 59848.47 &7.5 &N &Y &N &2.4693 &762.4 &5.2 &2.1 &6.2, 9.1\\
*J1843$-$0524t &18:43:47.6 &$-$05:24:35 & 59848.48 &7.5 &Y &N &N &0.0383 &256.8 &4.3 &0.0 &4.2, 5.2\\
J1844$-$0523t &18:44:14.3 &$-$05:23:44 & 59848.50 &7.5 &N &Y &N &1.2809 &405.2 &2.0 &8.0 &5.2, 6.4\\
*J1846$-$0512t &18:46:40.7 &$-$05:12:06 & 59426.66 &4.0 &Y &Y &Y &0.2753 &208.1 &29.6 &5.0 &4.2, 4.8\\
J1847$-$0624t &18:47:49.9 &$-$06:24:34 & 59848.53 &3.8 &N &Y &Y &0.1878 &388.2 &6.5 &0.2 &10.9, 7.5\\
J1854$-$0654t &18:54:00.9 &$-$06:54:19 & 59593.19 &17.5 &Y &Y &N &1.3806 &309.1 &7.6 &13.7 &16.8, 7.3\\
\hline
    \end{tabular}
\end{table*}

\subsection{Observations and Data Preparation}

We selected data from FAST Data Releases 1 through 23. Motivated by the expected Galactic pulsar distribution, in which pulsars are concentrated toward the Galactic plane \citep[e.g.,][]{Gullon+2014,Dirson+2022}, we restricted our sample to observations with declination $\text{Decl.}<-5\degr$ and Galactic latitudes $|b|<5\degr$. No FAST observations are available at Decl.$<-14\degr$ due to the telescope's zenith limit. The selected observations are shown in \autoref{fig:obs}. The resulting data set covers a total sky area of $\sim 3.6\deg^2$, with a total cumulative integration time of $\sim 500$ hr and a total data volume of $\sim 700$ TB. To avoid potential conflicts of interest, data from GPPS and CRAFTS, as well as from other projects recorded with the central beam (M01) were excluded. 

FAST observations were conducted with the $L$-band 19-beam receiver, centered at 1.25~GHz with a bandwidth of 500 MHz \citep{LiDi+2018,Jiang+2020}. The data were sampled at time resolutions of $49.152~\mu{\rm s}$, $98.304~\mu{\rm s}$, or $196.608~\mu{\rm s}$, and recorded in 8-bit PSRFITS files \citep{psrfits} with 4096, 2048, or 1024 frequency channels and four (AABBCRCI), two (AABB), or one (AA+BB) polarization parameters. The effective integrated times range from $\sim $ 2 min to $\sim 2~{\rm hr}$, depending on their original scientific purpose of each observation.

For data preprocessing, we used \texttt{digifil} command from \texttt{DSPSR} package\footnote{\url{https://dspsr.sourceforge.net}} \citep{vanStraten+2011} to sum the polarizations and convert the PSRFITS data into filterbank format \citep{filterbank} with 512 frequency channels. For data originally sampled at $49.152~\mu{\rm s}$, we downsampled to $98.304~\mu{\rm s}$ for subsequent pulsation searches. Although this downsampling slightly reduces sensitivity by broadening the effective pulse width, it is necessary given our computational and storage limitations. We then used the \texttt{filtool} command from \texttt{PulsarX}\footnote{\url{https://github.com/ypmen/PulsarX}} \citep{PulsarX} to concatenate the separated filterbank files, perform preliminary RFI mitigation, and optimize the bandpass. We also used \texttt{filtool} to excise the calibration signal injected parts, typically present at the beginning or the end of each observation of the data and split the observations conducted in ``ON-OFF'' mode into their respective segments. After these preprocessing steps, the total volume of data requiring pulsar searches was reduced to $\sim 9$ TB.

\subsection{Periodic Search}

We employed both FFT-based and FFA-based periodic searches. FFT-based periodic pulsation searches were performed using the PulsaR Exploration and Search TOolkit\footnote{\url{https://github.com/scottransom/presto}} \citep[\texttt{PRESTO},][]{Ransom+2011}. We first used the \texttt{PRESTO} routine \texttt{rfifind} to identify and excise corrupted frequency channels and time integrations affected by strong or periodic RFI. We then employed the GPU-accelerated versions of \texttt{prepsubband}\footnote{\url{https://github.com/zdj649150499/Presto_GPU}} and \texttt{prepsubband\_cu} (from \texttt{PrestoZL}, \footnote{\url{https://github.com/zhejianglab/PrestoZL}} \citealt{Mao+2025}) to generate barycentric, de-dispersed time series, writing the output to local solid-state drives (SSDs) to improve I/O performance. Dedispersion was carried out over a dispersion measure (DM) range of $0-3000~{\rm pc~cm^{-3}}$. The DM step sizes were computed using the \texttt{PRESTO} Python script \texttt{DDplan.py}, which selects intervals such that the dedispersion smearing between adjacent trials remains a small fraction of the effective time resolution, thereby ensuring an optimal balance between sensitivity and computational cost. The resulting DM step sizes were 0.1, 0.2, 0.3, 0.5, 1.0 and 3.0~${\rm pc~cm^{-3}}$ for DM ranges starting at 0, 37, 113.4, 189, 322.2, 544.2 and 1432.2~${\rm pc~cm^{-3}}$, respectively.

The de-dispersed time series were subsequently Fourier transformed for FFT-based periodic searches. The intermediate FFT files were stored temporarily on a Linux Tmpfs \footnote{\url{https://www.kernel.org/doc/html/latest/filesystems/tmpfs.html}} virtual disk, created in RAM to minimize I/O overhead. We used the \texttt{PRESTO} routine \texttt{accelsearch} to search for periodic signals, summing up to $h = 16$ harmonics and allowing a maximum drift parameter of $z_{\rm max} = 200$. Parallel execution of \texttt{accelsearch} jobs was managed using \texttt{GNU parallel}\footnote{\url{https://www.gnu.org/software/parallel}}. When GPUs were not occupied with dedispersion tasks, we additionally used the GPU-accelerated \footnote{Benchmarking on a system equipped with an NVIDIA A100--40G GPU and 20 Intel i7--12700K CPUs demonstrates that \texttt{accelsearch\_cu} achieves an end-to-end speedup of 56.4$\times$ relative to the CPU-based \texttt{PRESTO}'s \texttt{accelsearch} with \texttt{OpenMP} \citep{Mao+2025}.} implementation \texttt{accelsearch\_cu} provided by \texttt{PrestoZL} \citep{Mao+2025}, which produces outputs consistent with the standard \texttt{accelsearch}. Candidates were sifted using \texttt{PRESTO} Python script \texttt{ACCEL\_sift.py} with a significance threshold of sigma$>4.0$. Promising candidates were then folded using \texttt{PRESTO}'s \texttt{prepfold} command, and the resulting diagnostic plots were examined visually.

In addition to FFT-based searches, we applied a phase-coherent search technique based on the FFA. Unlike Fourier methods, the FFA operates entirely in the time domain and provides enhanced sensitivity to long-period or intermittently emitting pulsars \citep{Cameron+2017,Morello+2020}. The FFA search was performed using the \texttt{RIPTIDE} package\footnote{\url{https://github.com/v-morello/riptide}} \citep{Morello+2020}. The \texttt{rffa} pipeline was applied to the time-series files produced by \texttt{prepsubband}, adopting a signal-to-noise ratio threshold of 8.0 and searching trial periods from 0.1 to 30~s. Periods below 0.1~s were excluded due to prohibitive computational cost. Diagnostic plots generated by \texttt{rffa} were inspected visually, and promising candidates were subsequently folded with \texttt{prepfold} to assess their astrophysical nature or identify potential RFI. The criteria used to distinguish astrophysical signals from RFI for both periodic and single-pulse (see below) signals are described in detail in the appendix of \citet{Gao+2025b}.

\subsection{Single-pulse Search}

Single-pulse searches were performed using \texttt{TransientX}\footnote{\url{https://github.com/ypmen/TransientX}} \citep{TransientX}, a high-performance CPU-based single-pulse search package. We employed the same dedispersion grid used in periodic search. The maximum pulse width was set to the default value of $50~{\rm ms}$, and a signal-to-noise ratio threshold of 7.0 was applied. Diagnostic plots, including the dedispersed pulse profile, frequency-time waterfall, and DM-time space, were produced using the single-pulse analysis toolkit \texttt{YOUR}\footnote{\url{https://github.com/thepetabyteproject/your}} \citep{Aggarwal+2020}. Given the large number of detected candidates, we used \texttt{FETCH}\footnote{\url{https://github.com/devanshkv/fetch}} \citep{FETCH}, an open-source convolutional neural network-based classifier, to automatically evaluate the diagnostic plots. A candidate was retained if at least one of the five trained models identified it as promising. The reduced set of candidates was then inspected visually to identify the astrophysical single pulses.


\section{Results and Discussion}\label{sec:res}

\subsection{Discovery of 19 New Pulsars}
\begin{figure*}
    \centering
\centering
\setlength{\tabcolsep}{2pt}
\begin{tabular}{ccccc}
\includegraphics[width=0.18\textwidth,height=0.29\textwidth]{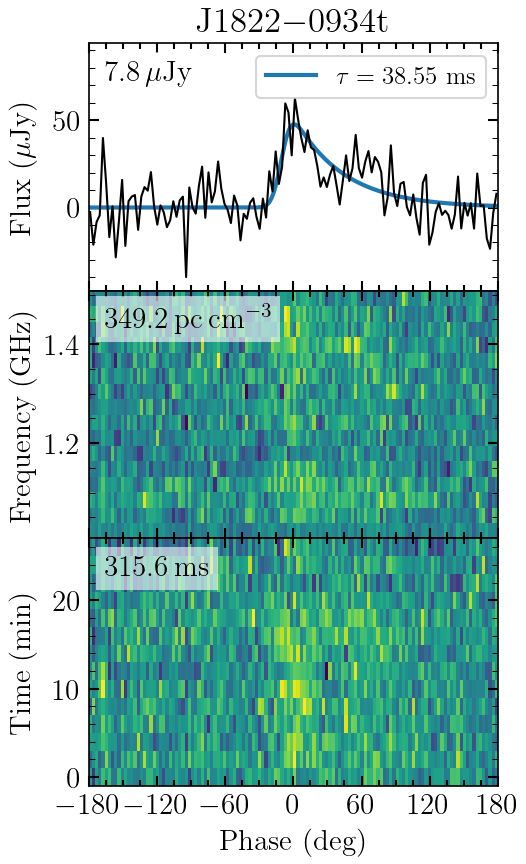}&
\includegraphics[width=0.18\textwidth,height=0.29\textwidth]{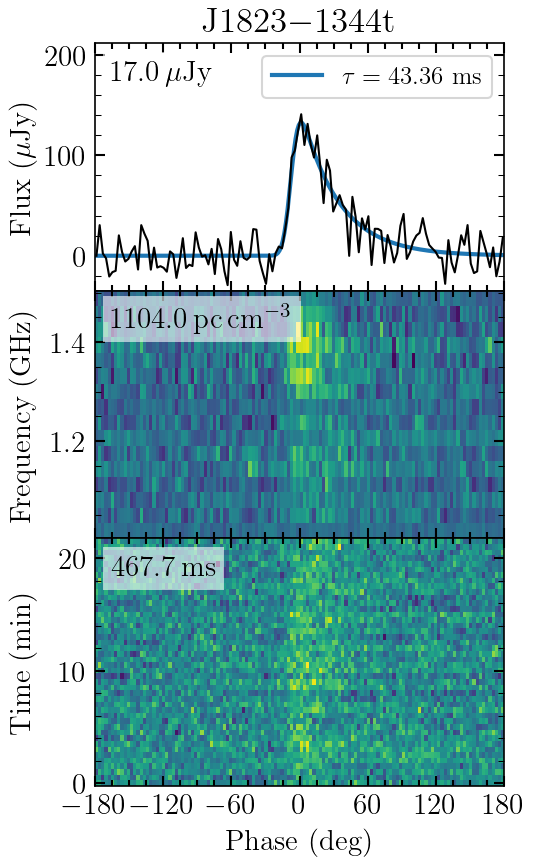}&
\includegraphics[width=0.18\textwidth,height=0.29\textwidth]{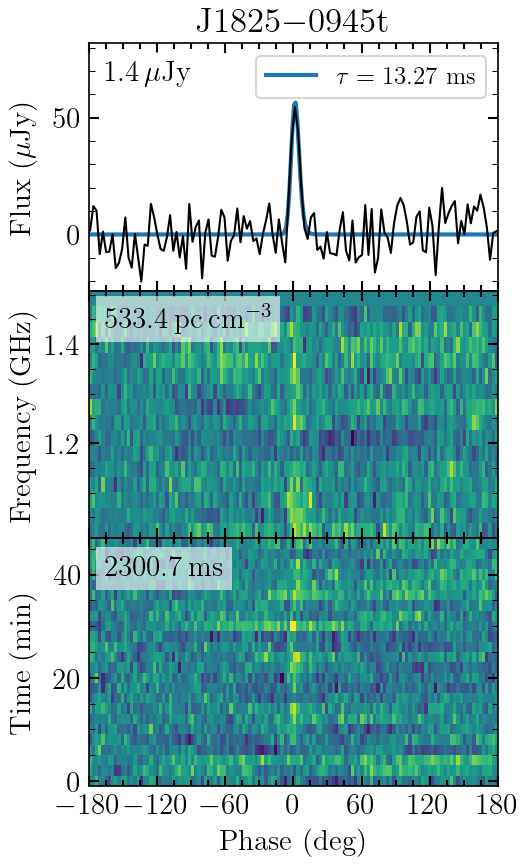}&
\includegraphics[width=0.18\textwidth,height=0.29\textwidth]{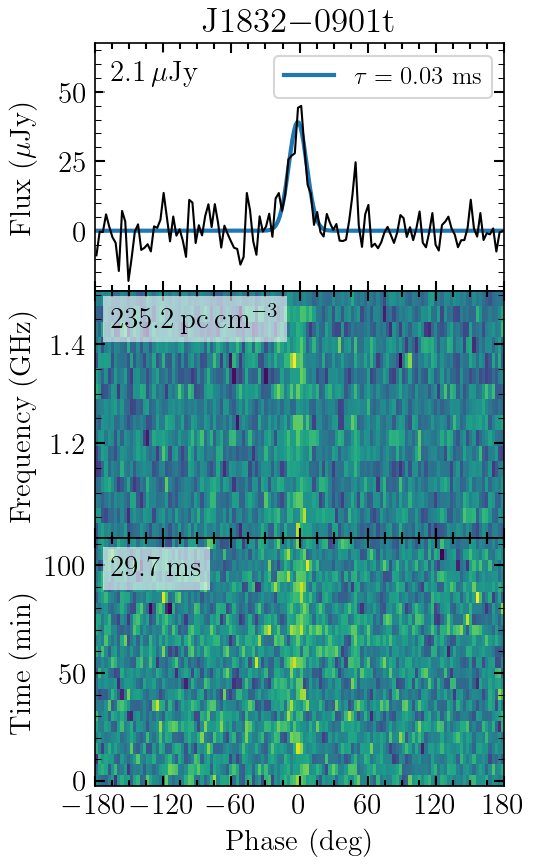}&
\includegraphics[width=0.18\textwidth,height=0.29\textwidth]{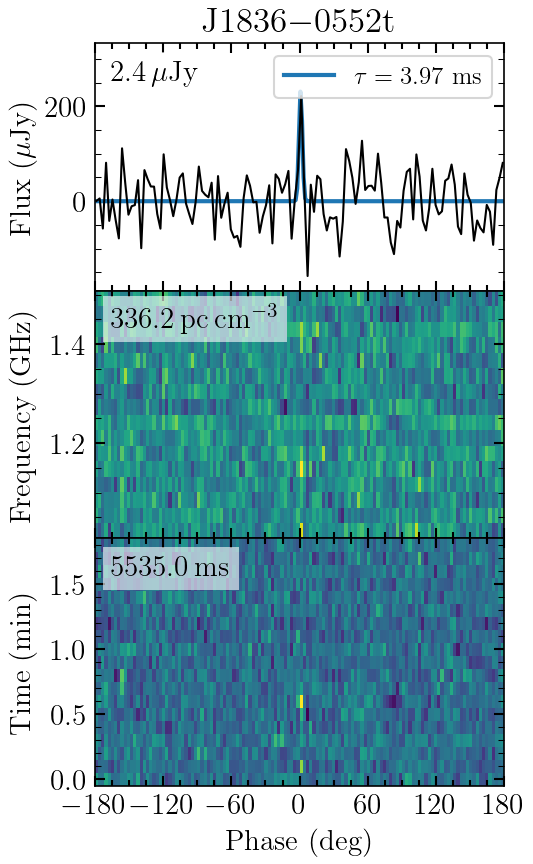}\\
\includegraphics[width=0.18\textwidth,height=0.29\textwidth]{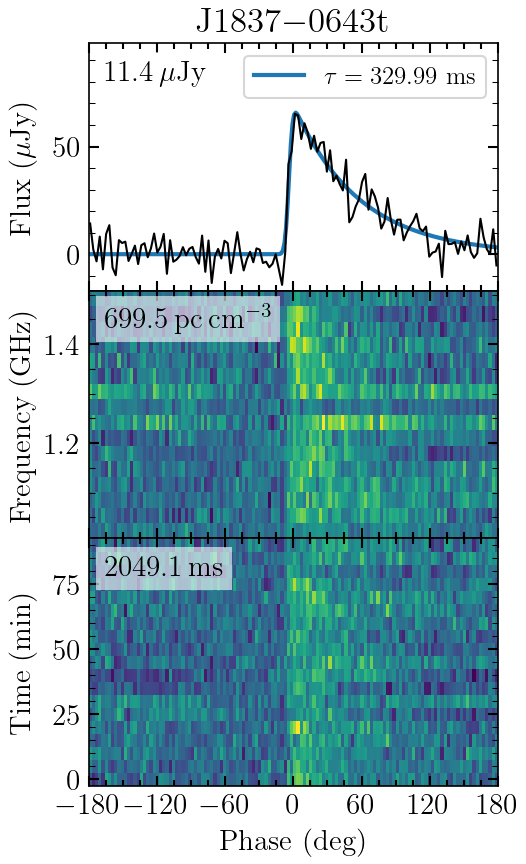}&
\includegraphics[width=0.18\textwidth,height=0.29\textwidth]{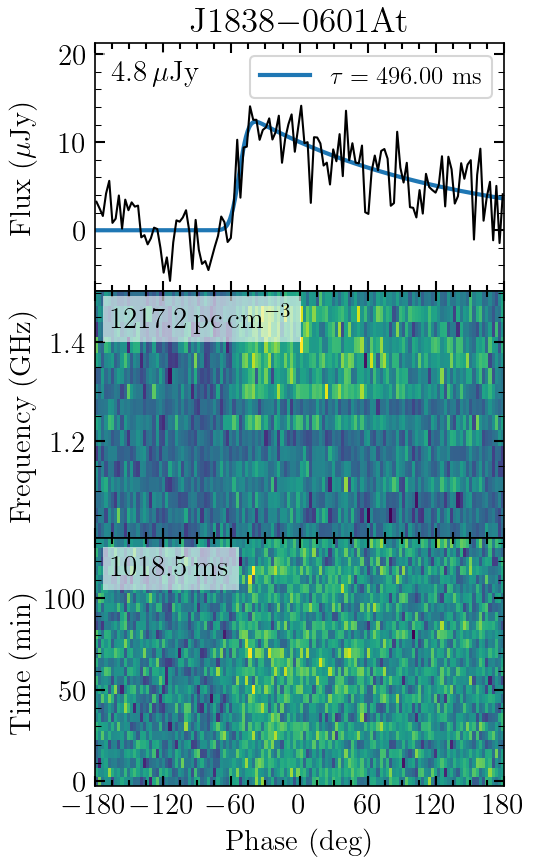}&
\includegraphics[width=0.18\textwidth,height=0.29\textwidth]{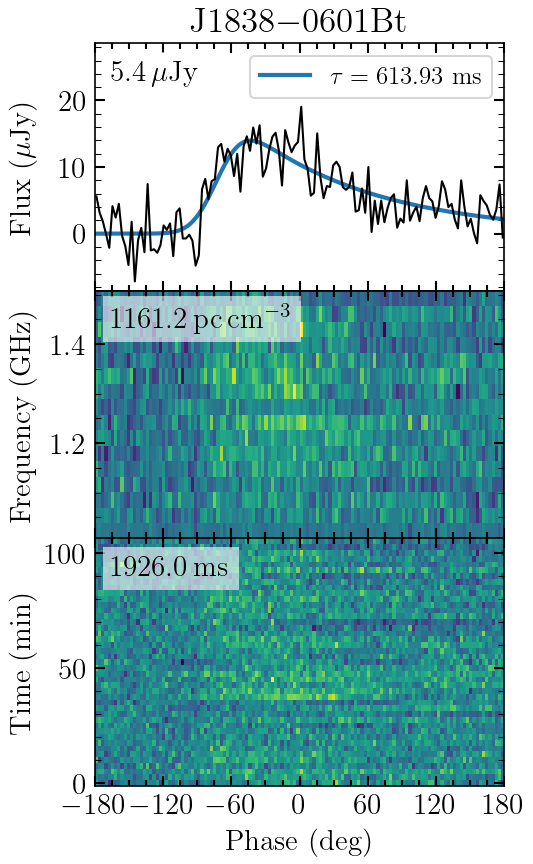}&
\includegraphics[width=0.18\textwidth,height=0.29\textwidth]{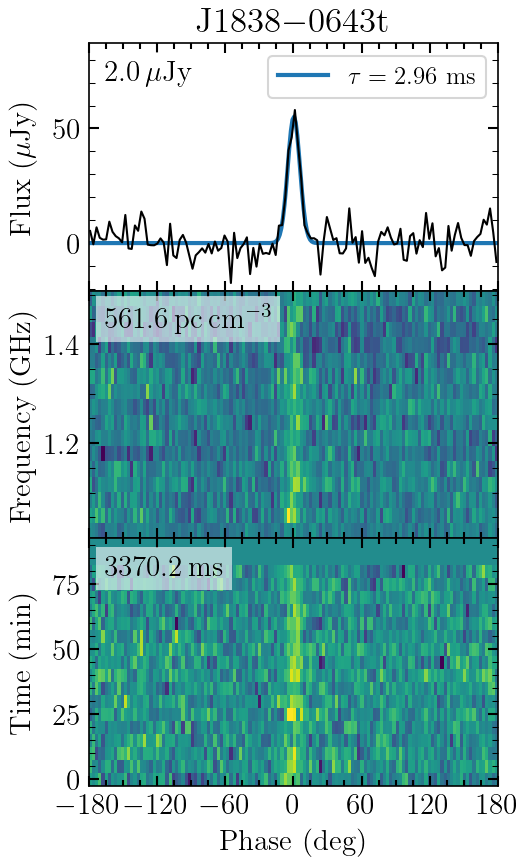}&
\includegraphics[width=0.18\textwidth,height=0.29\textwidth]{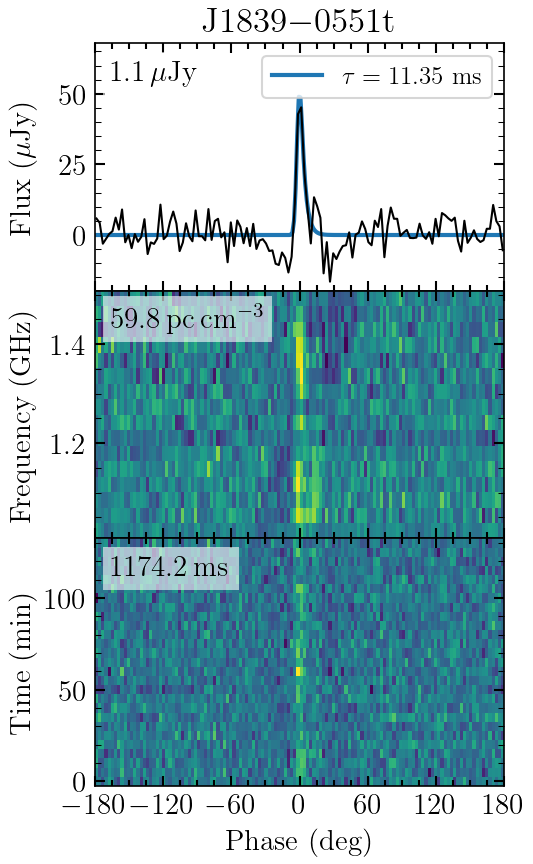}\\
\includegraphics[width=0.18\textwidth,height=0.29\textwidth]{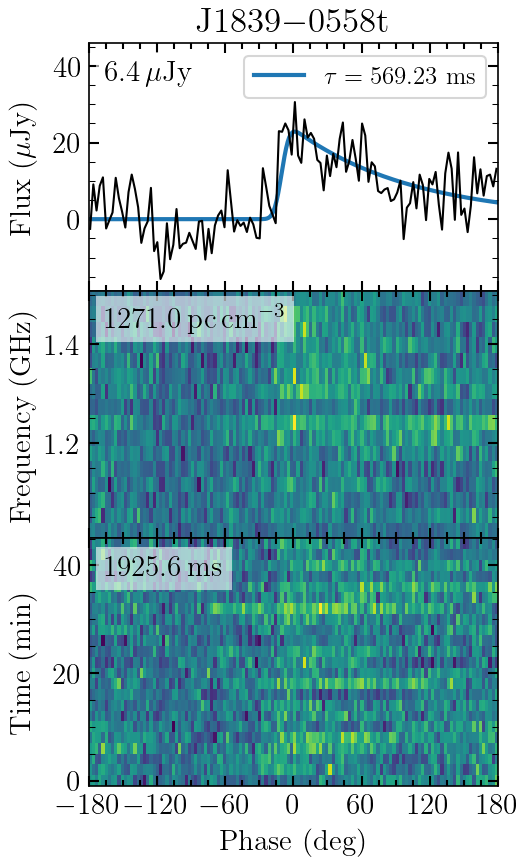}&
\includegraphics[width=0.18\textwidth,height=0.29\textwidth]{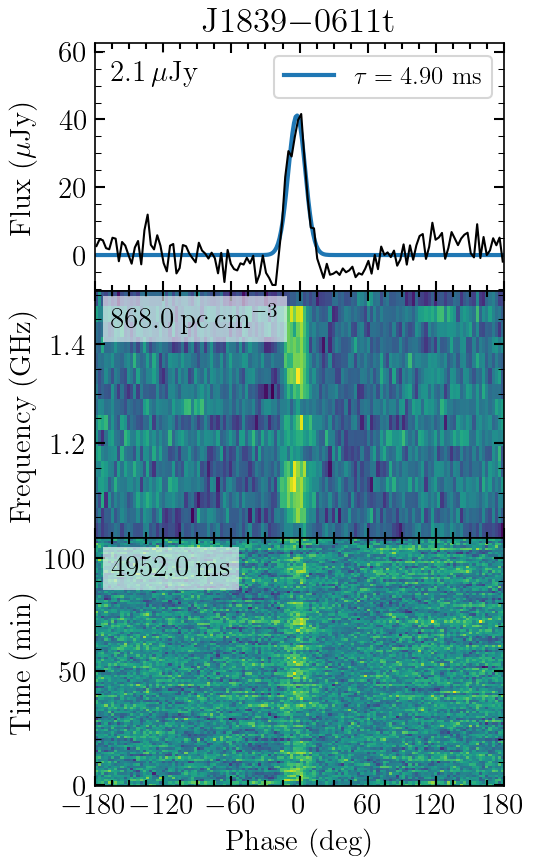}&
\includegraphics[width=0.18\textwidth,height=0.29\textwidth]{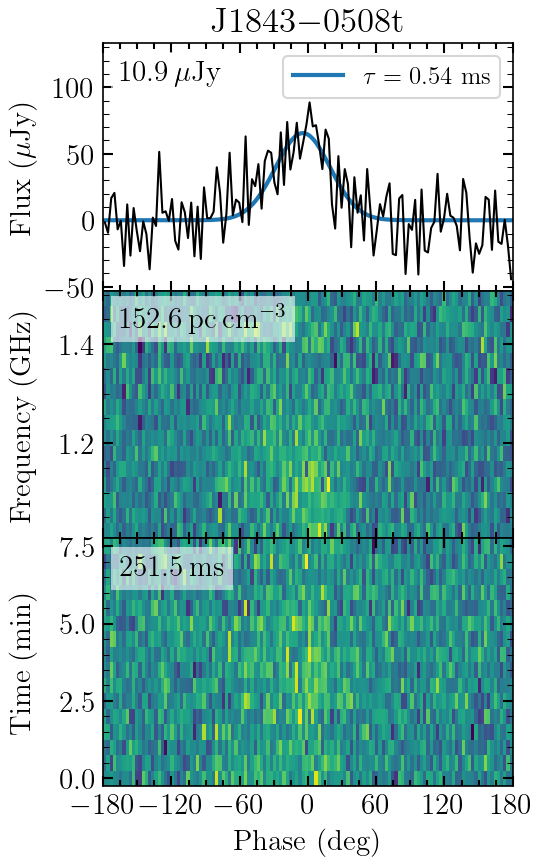}&
\includegraphics[width=0.18\textwidth,height=0.29\textwidth]{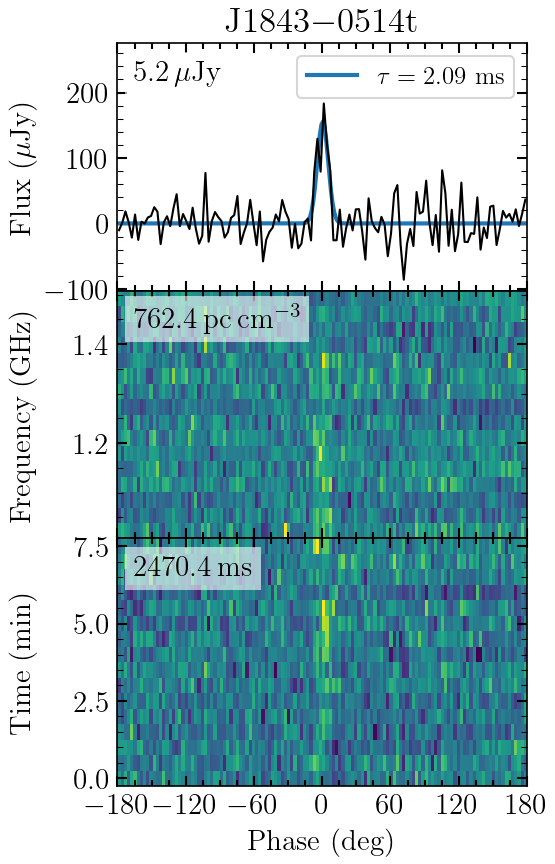}&
\includegraphics[width=0.18\textwidth,height=0.29\textwidth]{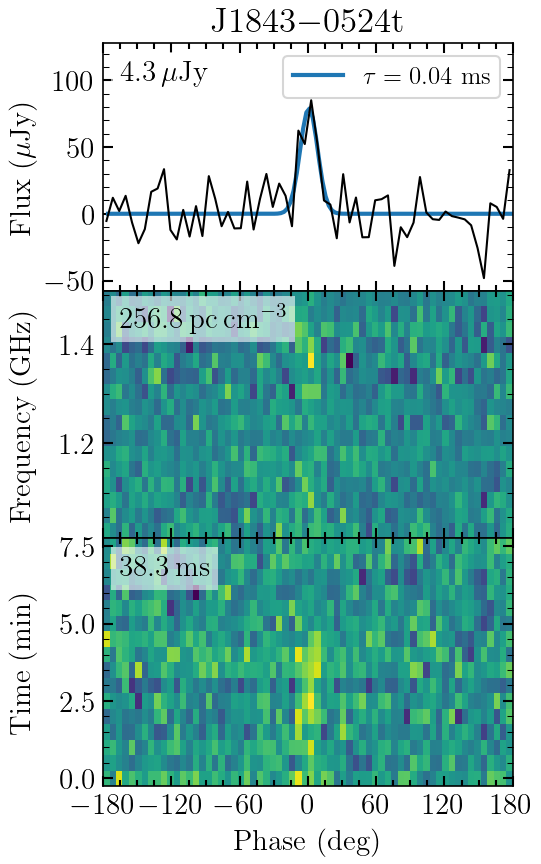}\\
\includegraphics[width=0.18\textwidth,height=0.29\textwidth]{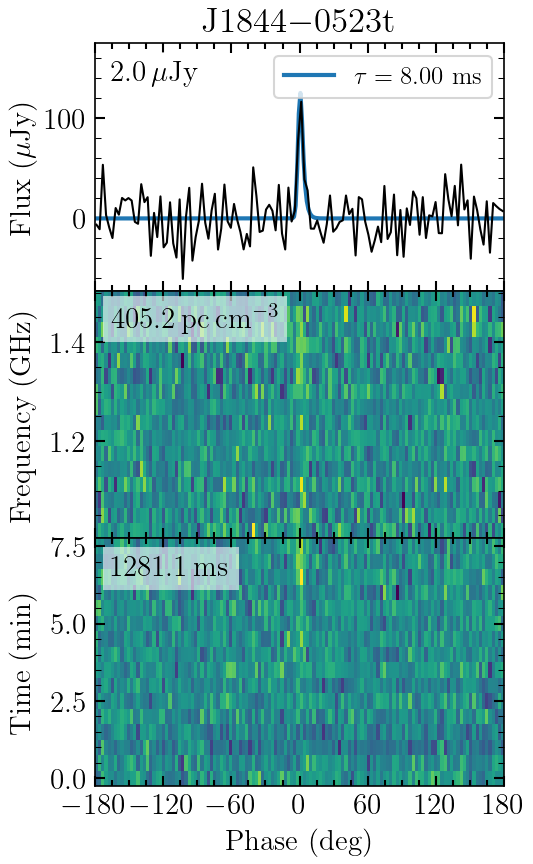}&
\includegraphics[width=0.18\textwidth,height=0.29\textwidth]{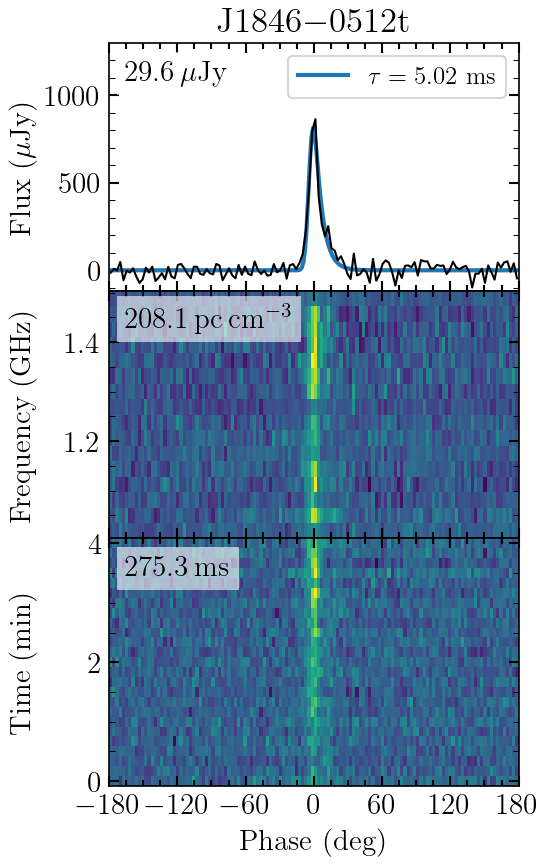}&
\includegraphics[width=0.18\textwidth,height=0.29\textwidth]{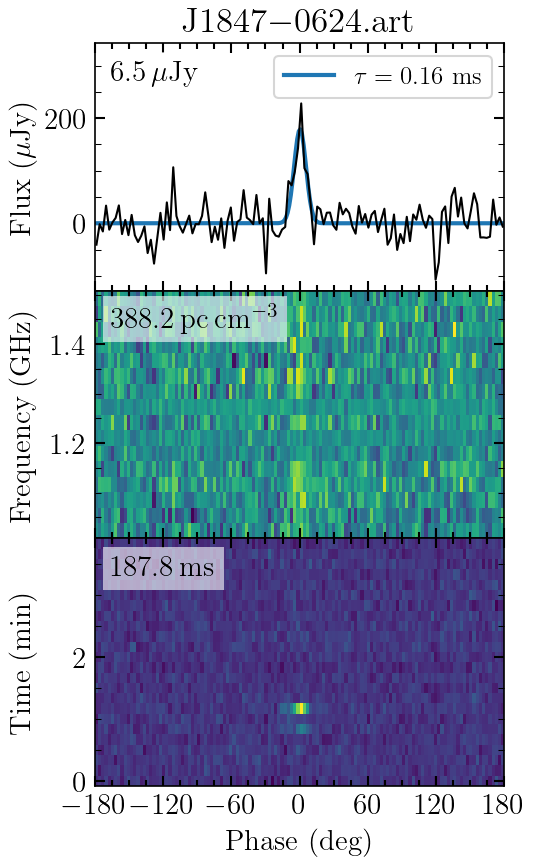}&
\includegraphics[width=0.18\textwidth,height=0.29\textwidth]{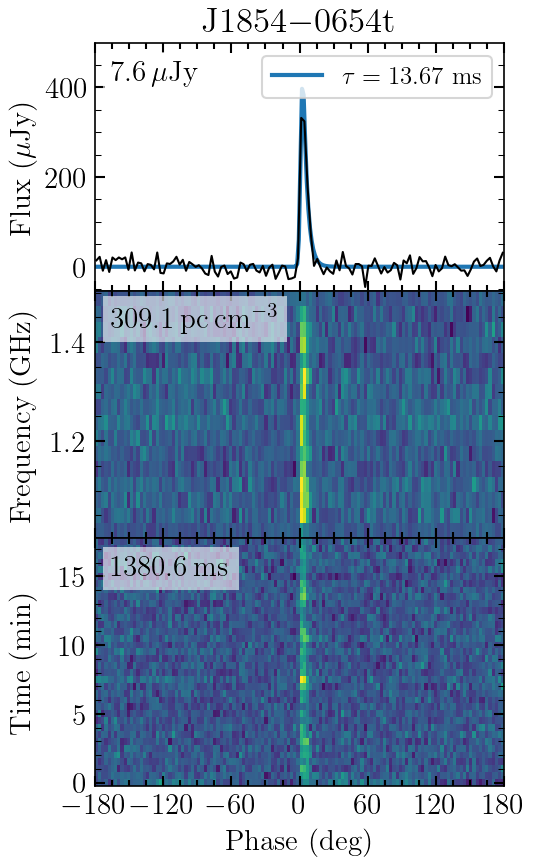}&
\multicolumn{1}{c}{} \\
\end{tabular}
\caption{Diagnostic plots for the 19 newly discovered pulsars reported in this work. For each pulsar, the top panel shows the integrated pulse profile together with the best-fitting scattering tail (solid blue curve, \autoref{eq:profile}). The middle panel presents the dedispersed, folded frequency-phase diagram. The bottom panel displays the time-phase diagram over the full observing span.\label{fig:psrs}}
\end{figure*}

\begin{figure}
    \centering
    \includegraphics[width=\linewidth]{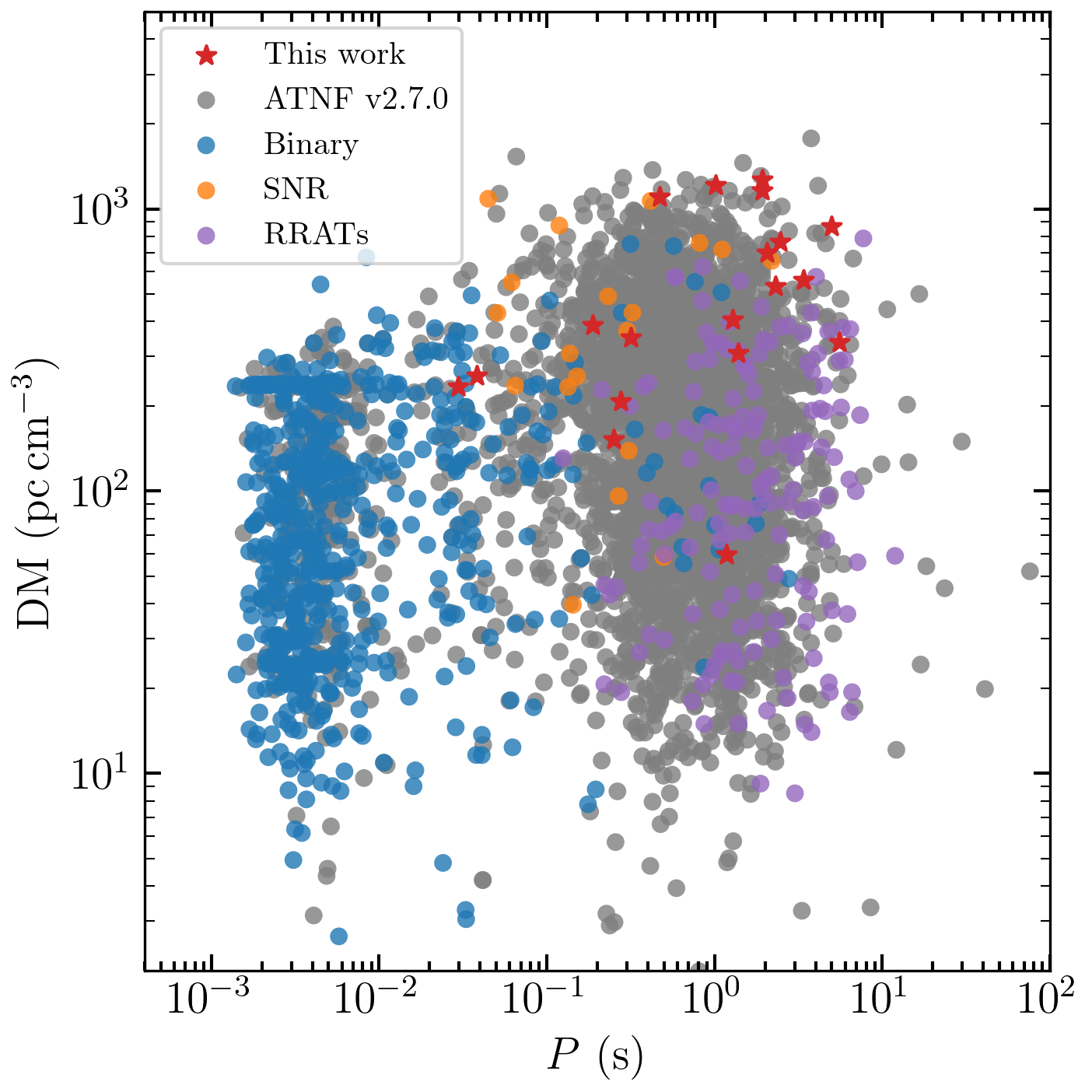}
    \caption{Distribution of the newly discovered pulsars (red stars) in the spin period-DM plane. Circles show all known pulsars listed in the ATNF pulsar catalogue \citep[v2.7.0,][]{ATNF}. Pulsars in binary systems, supernova-remnant (SNR) associations, and rotating radio transients (RRATs) from the catalogue are highlighted in blue, orange, and purple, respectively.\label{fig:dmp}}
\end{figure}

We used the Pulsar Survey Scraper\footnote{\url{https://pulsar.cgca-hub.org/search}} \citep{Kaplan+2022} to check whether the pulsars we discovered have already been reported by other surveys. The Pulsar Survey Scraper gathers newly reported pulsar discoveries before they appear in the ATNF Pulsar Catalogue and allows searching and filtering based on sky position and DM. After a careful inspection, we identified 19 new pulsars\footnote{These pulsar discoveries are available at \url{https://github.com/astro-sjgao/PulsarGleaners}. Just before submitting this paper, we re-checked using the Pulsar Survey Scraper and found that J1843$-$0524t had been discovered by the GPPS survey on 2025-10-11, based on data observed on 2025-09-26, with the information posted online on 2025-11-03. We note that we independently discovered this pulsar on 2025-10-06 using archived data.}. The basic parameters of the 19 newly discovered pulsars are listed in \autoref{tab:psrs}. Among these, 12 were detected using FFT-based search, while 17 were detected using FFA-based search (see \autoref{tab:psrs}). The two pulsars not captured by FFA have spin periods below the lower bound of our adopted FFA search range (0.1$-$30~s). This result highlights the sensitivity and effectiveness of the FFA-based method within its operational parameter space. The reported topocentric spin period and DM values were refined by re-folding the full-resolution raw data with \texttt{psrfold\_fil2} from \texttt{PulsarX} \citep{PulsarX}. For each source, we also provide the R.A. and Decl. of the beam center in which the pulsar was detected. Given that the FAST 19-beam receiver has a beam width of $\sim 3'$ and a pointing accuracy of $7''.9$ \citep{Jiang+2020}, the true sky position of each pulsar is expected to be uncertain by about $1'.5$ relative to the beam center. To indicate this positional uncertainty, we append the suffix ``t'' (for ``temporary'') to their provisional pulsar names. The integrated pulse profiles, dedispersed frequency-phase diagrams, and time-phase diagrams for all 19 pulsars are presented in \autoref{fig:psrs}, while their spin periods versus DM values are shown in \autoref{fig:dmp}.

The spin periods of the newly discovered pulsars span a broad range from 0.03 to 5.54~s. As shown in \autoref{fig:dmp}, only two pulsars (PSRs~J1832$-$0901t and J1843$-$0524t) have periods $<0.1~{\rm s}$. These lie in a region overlapping young, non-recycled pulsars and mildly recycled systems. The former are often associated with supernova remnants \citep[e.g., the Crab pulsar,][]{Staelin+1968}, while recycled pulsars may have massive white dwarf or neutron star companions \citep[e.g.,][]{Berezina+2017}. For PSR~J1832$-$0901t, the discovery observation shows no significant acceleration, whereas the confirmation observation, due to its shorter integration time, is not recovered in a blind search and can be folded only using the parameters obtain from the discovery observation. PSR~J1843$-$0524t behaves similarly. Additional timing is needed to determine if either is in a binary. J1839$-$0558t has the largest DM value in our sample, $\sim 1271~{\rm pc~cm^{-3}}$, second only to the most highly dispersed FAST-discovered pulsar, PSR~J1843$-$0310g \citep[1290~${\rm pc~cm^{-3}}$,][]{Han+2025}. In \autoref{tab:psrs}, we listed the distance estimates based on the YMW16 \citep{YMW16} and NE2001 \citep{Cordes+2002} electron density models. PSRs~J1847$-$0624t and J1836$-$0552t exhibit sporadic emission (see \autoref{fig:psrs}) consistent with the characteristic behavior of RRATs \citep[][]{McLaughlin+2006}.

In the following sections, we present the scattering properties (Section~\ref{sec:scattering}), flux-density estimates and search sensitivities of the newly discovered pulsars (Section~\ref{sec:flux}), and then describe the two RRAT-like sources and the effectiveness of FFA in revealing faint and sporadic pulsars in more detail (Section~\ref{sec:rrat}).

\subsection{Pulsar Scattering}\label{sec:scattering}

\begin{figure}
    \centering
    \includegraphics[width=\linewidth]{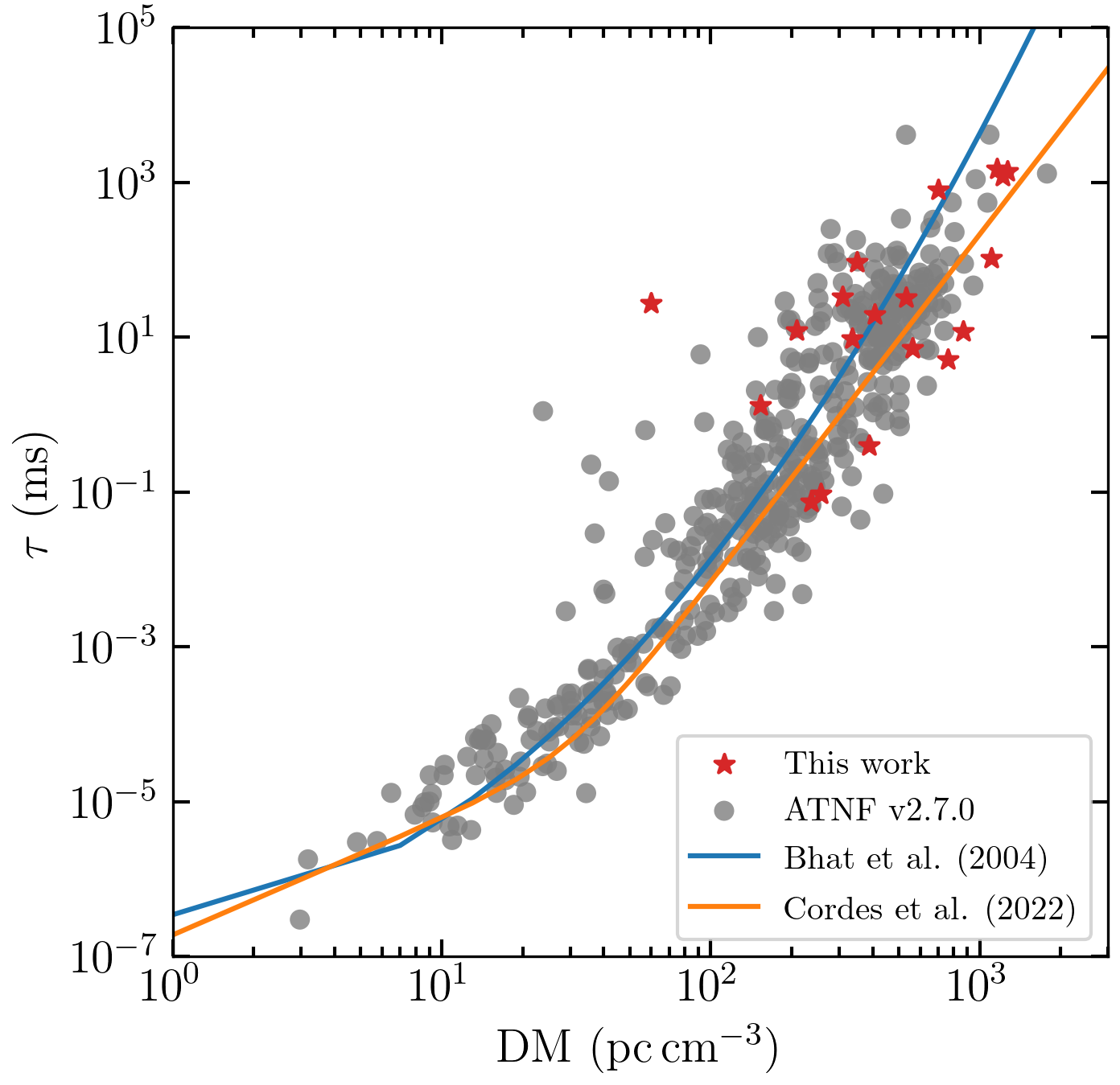}
    \caption{Scatter-broadening timescale $\tau$ vs. DM ($\nu=1~{\rm GHz}$). The newly discovered pulsars are shown as red stars, and pulsars from the ATNF Pulsar Catalogue are shown as gray circles. The empirical $\tau$-DM relations from \cite{Bhat+2004} and \cite{Cordes+2022} are plotted as blue and orange curves, respectively.\label{fig:tau}}
\end{figure}

As shown in \autoref{fig:psrs}, several pulsars exhibit significant scattering broadening, e.g., PSRs~J1823$-$1345t and J1837$-$0643, caused by small-scale irregularities in the ionized interstellar medium along the propagation path \citep{Rickett+1977}. Scattering asymmetrically broadens the intrinsic narrow pulse. A commonly used approximation models the scattered pulse profile as the convolution of the intrinsic Gaussian-like pulse shape with an exponential pulse-broadening function corresponding to isotropic scattering in a thin screen \citep{Cordes+2001}. The resulting profile can be expressed analytically as an exponentially modified Gaussian \citep[e.g.,][]{Jankowski+2023}:  
\begin{equation}\label{eq:profile}
\begin{aligned}
    f(t)={}&b+\frac{F}{2\tau}\exp\left(\frac{\sigma^2}{2\tau^2}-\frac{t-\mu}{\tau}\right)\\
    &\times{\rm erfc}\left[-\frac{1}{\sqrt2}\left(\frac{t-\mu}{\sigma}-\frac{\sigma}{\tau}\right)\right]
\end{aligned}
\end{equation}
where $t$ is the time, $b\simeq 0$ is the baseline offset, $F$ is the pulse fluence (total pulse energy) of the intrinsic Gaussian component, $\tau$ is the scatter-broadening timescale, and $\mu$ and $\sigma$ describe the location and width of the intrinsic Gaussian component. Because the newly discovered pulsars are faint, reliable sub-band scattering fits are difficult to obtain. Therefore, we fit the frequency-summed pulse profiles using \autoref{eq:profile}. The resulting scatter-broadening timescales $\tau$ are listed in \autoref{tab:psrs}, and the corresponding best-fit profiles are shown in \autoref{fig:psrs}.

The pulse-broadening time $\tau$ correlates with DM and depends strongly on observing frequency. For example, \cite{Bhat+2004} provides
\begin{equation}\label{eq:tauDM2004}
\begin{aligned}
\log\left(\frac{\tau}{\rm ms}\right)={}&-6.46+0.154\log\left(\frac{\text{DM}}{\rm pc~cm^{-3}}\right)\\&+1.07\log\left(\frac{\text{DM}}{\rm pc~cm^{-3}}\right)^2\\&-3.86\log \left(\frac{\nu}{\rm GHz}\right)
\end{aligned}
\end{equation}
with $\nu$ is the center frequency, and \cite{Cordes+2022} give
\begin{equation}\label{eq:tauDM2022}
\begin{aligned}
    \tau={}&1.9\times10^{-7}~{\rm ms}\times \left(\frac{\text{DM}}{\rm pc~cm^{-3}}\right)^{1.5}\\&\times\left[1+3.55\times10^{-5}\left(\frac{\text{DM}}{\rm pc~cm^{-3}}\right)^3\right],
\end{aligned}
\end{equation}
with the reference frequency at 1~GHz. We scale our measured $\tau$ values from center frequency $\nu=1.25~{\rm GHz}$ to 1~GHz using $\tau \propto \nu^{-4}$ and compare them with the predictions of \autoref{eq:tauDM2004} and \autoref{eq:tauDM2022} in \autoref{fig:tau}. The $\tau$ values of the newly discovered pulsars are broadly consistent with empirical relations. Our discoveries contribute new data points at the high-DM end of this relation.

\subsection{Search Sensitivities}\label{sec:flux}
\begin{figure}
    \centering
    \includegraphics[width=\linewidth]{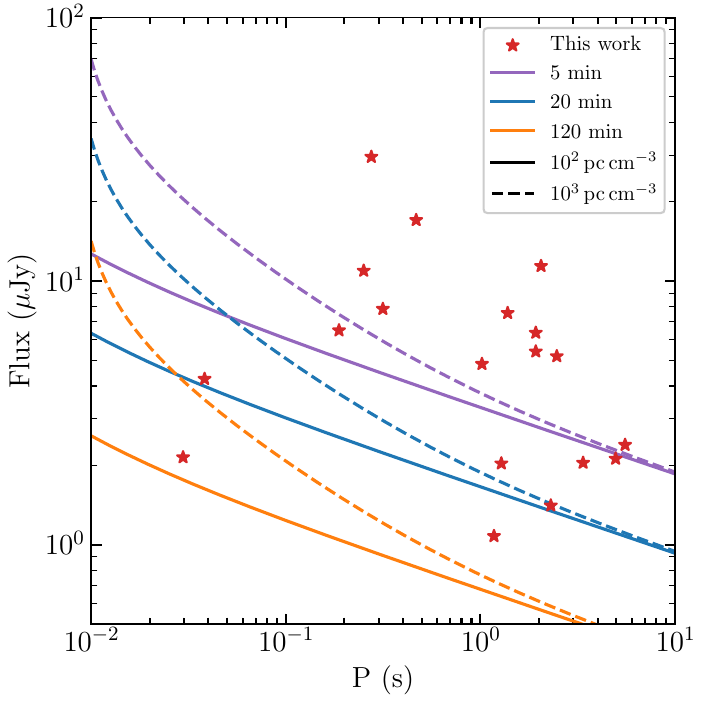}
    \caption{Search sensitivity curves for different integration times and DM values. Purple, blue, and orange lines correspond to integration times of 5 min, 20 min, and 2 hr, respectively. Solid and dashed lines show sensitivities for DM$=10^{2}$ and $10^{3}~\rm{pc~cm^{-3}}$, respectively. The red stars mark the estimated mean flux densities of the newly discovered pulsars.\label{fig:sn}}
\end{figure}
The rms variation in the system noise is described by the radiometer equation \citep[e.g.,][]{hpa},
\begin{equation}\label{eq:Ssys}
    S_{\rm sys}=\frac{T_{\rm sys}}{G\sqrt{N_{\rm p}\delta t\delta \nu}}=C\sigma_{\rm p}
\end{equation}
where $T_{\rm sys}=24~{\rm K}$ is the typical system temperature, $G=16~{\rm K}$ is the telescope gain, $N_{\rm p}=2$ is the number of summed polarizations, $\delta t$ is the integration time per phase bin, and $\delta \nu$ is the total bandwidth ($\sim 75\%$ of the $500~{\rm MHz}$ total bandwidth was used after RFI migration and bad edge removal). The true pulsar flux density from the summed pulse profiles is obtained by scaling the off-pulse rms $\sigma_{\rm p}$ by the factor $C$. This flux estimate should be regarded as a lower limit for the following reasons: (1) the telescope gain is reduced at zenith angles $\gtrsim26\degr$, where only part of the FAST's 300-m effectively reflector is illuminated \citep{Jiang+2020}; (2) the true pulsar position may be offset from the beam center, further reducing effective gain \citep[e.g.,][]{Ng+2015}; (3) the system temperature of FAST increases with increasing zenith angle due to emission of a nearby mountain through the sidelobe when zenith angle $\gtrsim 15\degr$ \citep{Jiang+2020} and (4) contributions from the sky background to the system temperature are not included. The mean flux densities of the discovered pulsars are calculated using the fluence derived from pulse profile fitting (\autoref{eq:profile}) and divided by the spin period, which are listed in \autoref{tab:psrs}.

The search sensitivities $S_{\rm min}$ of periodic signals for a given integration time is \citep[e.g.,][]{hpa}
\begin{equation}
    S_{\rm min}={\rm S/N}_{\rm min}\times\frac{T_{\rm sys}}{G\sqrt{N_{\rm p} \delta \nu T_{\rm obs}}}\sqrt{\frac{W_{\rm eff}}{P-W_{\rm eff}}}, 
\end{equation}
where ${\rm S/N}_{\rm min}=10$ is the desired signal-to-noise ratio threshold for the periodic search, $T_{\rm obs}$ is the total observation duration. The parameters $P$ and $W_{\rm eff}$ represent the pulsar period and the effective pulse width, respectively. The effective pulse width is given by \begin{equation}
    W_{\rm eff}=\sqrt{W_{\rm i}^2+W_{\rm s}^2+\tau^2+ W_{\rm DM}^2},
\end{equation}
where $W_{\rm i}$ is the intrinsic pulse width, and is assumed 10\% of the pulse period for $P<10~{\rm ms}$ and reduced with $1/\sqrt{P}$ for $P>10~{\rm ms}$ \citep[e.g.,][]{Han+2021}. $W_{\rm s}$ is the sampling time (typically $98.304~{\mu\rm s}$), $\tau$ is the scatter-broadening time (\autoref{eq:tauDM2022} is applied), and
\begin{equation}
\begin{aligned}
    W_{\rm DM}={}&8.3\times10^{-3}~{\rm ms}\\
    &\times\left(\frac{\text{DM}}{\rm pc~cm^{-3}}\right)\left(\frac{\delta\nu/N_{\rm c}}{\rm MHz}\right)\left(\frac{\nu}{\rm GHz}\right)^{-3}
\end{aligned}
\end{equation}
is the uncorrected dispersion delay within a frequency channel with $N_{\rm c}=512$ the total number of frequency channels. Here $\nu=1.0~{\rm GHz}$ is applied rather than center frequency of 1.25~GHz.

\autoref{fig:sn} presents the search sensitivities for DM values of $100~{\rm pc~cm^{-3}}$ and $1000~{\rm pc~cm^{-3}}$, and for integration times of 5~min, 20~min and 2~hr. The estimated mean flux densities of the pulsars are marked with red stars. For a typical 5-min integration, only a subset ($\gtrsim 50\%$) of the pulsars are detectable, which corresponds to the typical sensitivity of GPPS \citep{Han+2021}. For FAST's CRAFTS pulsar survey \citep{LiDi+2018}, which is conducted in drift-scan mode, the effective integration time for a single beam is only $\sim 12~{\rm s}$ at $\rm Decl.=-5\degr$, resulting in substantially reduced sensitivity ($1/5$ of that achieved with a 5-min integration). As shown in \autoref{fig:sn}, about half of our pulsars require a much longer integration time ($>20~{\rm min}$) to be detected. This suggests that some faint pulsars remain still undetected in previous FAST pulsar surveys. 

In \autoref{fig:sn}, the plotted fluxes represent mean flux densities. Pulsars with sporadic emission may still be detectable in single-pulse searches, but they appear to have extremely low mean fluxes due to nulling \citep[see also,][]{Han+2021,Zhou+2023,Han+2025}. The sensitivity for single-pulse searches can be estimated using \autoref{eq:Ssys} by replacing $\delta t$ with the desired effective pulse width and multiplying by the target signal-to-noise ratio ${\rm S/N}_{\rm min}$. For a minimum ${\rm S/N}_{\rm min}=7$ and $\delta t=1~{\rm ms}$, a sensitivity of $\sim 10~{\rm mJy}$ can be achieved in our single-pulse searches.

\begin{figure*}
    \centering
    \plottwo{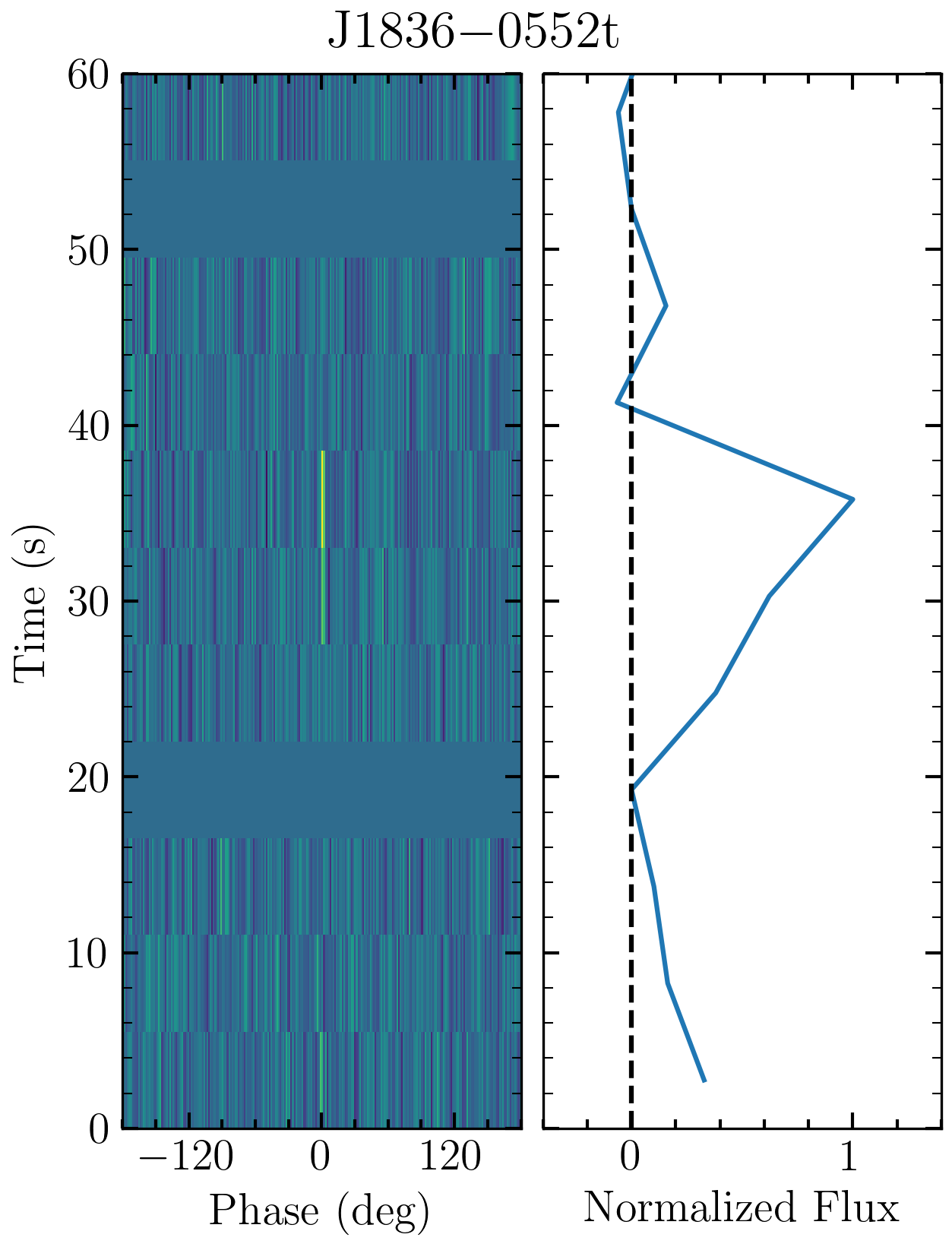}{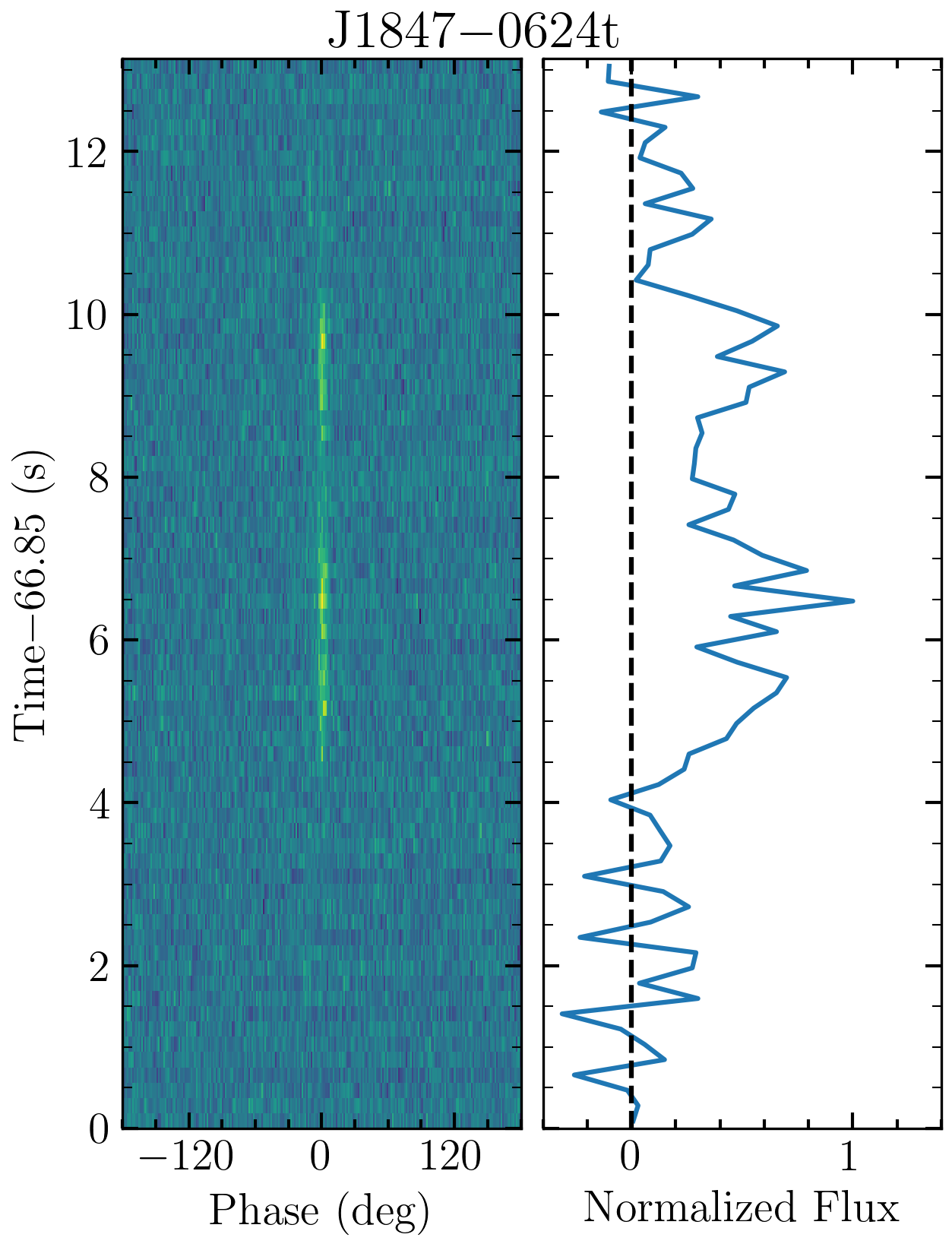}
    \caption{Single-pulse plots for PSRs~J1836$-$0552t (left) and J1847$-$0624t (right), showing their sporadic emission in the phase-time plane and phase-normalized flux plane.\label{fig:sp}}
\end{figure*}

\subsection{Discovery of Two Rotating Radio Transients}\label{sec:rrat}

RRATs are a distinctive subclass of pulsars characterized by irregular, sporadic single pulses \citep{McLaughlin+2006}. Observationally, they are defined as sources that are more easily detected via single-pulse searches than in periodic search \citep[e.g.,][]{Keane+2011,Zhou+2023}. A physically motivated interpretation describes RRATs as pulsars with extremely high nulling fractions \citep[e.g.,][]{Burke-Spolaor+2010,Gao+2025}. We identified two such pulsars exhibiting sporadic emission, PSRs~J1836$-$0552t and J1847$-$0624t. PSR~J1836$-$0552t was discovered through both single-pulse and FFA searches in a 2-min observation. The single-pulse search detected three significant pulses at ${\rm DM} \simeq 340~{\rm pc~cm^{-3}}$, while the FFA search yielded a period of $\simeq 5.5~{\rm s}$ and ${\rm DM} \simeq 346~{\rm pc~cm^{-3}}$. PSR~J1847$-$0624t was similarly discovered using single-pulse and FFA techniques in a 5-min observation.

We used \texttt{DSPSR} to fold both PSR~J1836$-$0552t and PSR~J1847$-$0624t in single-pulse mode. The left panel of \autoref{fig:sp} shows the pulse phase versus time for the first 1-min observation of PSR~J1836$-$0552t, with the normalized flux density of each detected pulse displayed on the right side. Only two significant pulses are present (during the 6th and 7th rotation periods) consistent with the results of the single-pulse search. The right panel of \autoref{fig:sp} shows the pulse sequence of PSR~J1847$-$0624t, where the first integration begins at 66.85~s. Emission becomes visible at $\sim 71$~s and persists for only $\sim 6$~s. We estimate that $\sim 86\%$ and $\sim 97\%$ of rotations are null for PSRs~J1836$-$0552t and J1847$-$0624t, respectively. Longer and dedicated follow-up observations will be essential to better characterize the emission properties and underlying nature of these two RRATs.

Notably, our FFT-based periodic search failed to detect either of these RRAT-like pulsars. In addition, three faint pulsars (PSRs~J1825$-$0945t, J1843$-$0514t and J1844$-$0523t; see \autoref{tab:psrs}) were also missed by the standard FFT-based pipeline due to their low flux densities. For long-period pulsars, FFT-based searches sum only a limited number of harmonics (usually 8--32) due to the computational costs, resulting in the loss of signal power. In contrast, the FFA folds the time-series directly, providing enhanced sensitivity to long-period and highly intermittent sources. These non-detections underscore the effectiveness of FFA and single-pulse techniques for uncovering long-period, sporadic and faint pulsars.

\section{Conclusion}\label{sec:sum}

We conducted a systematic search of FAST archival observations covering $|b|<5\degr$ and Decl.$<-5\degr$. This led to the discovery of 19 previously unreported pulsars. The spin periods of these pulsars span a broad range from 0.03 to 5.54~s. Only two pulsars, PSRs~J1832$-$0901t and J1843$-$0524t, have periods shorter than $0.1~{\rm s}$, which places them in the regime of either young, non-recycled pulsars or mildly recycled binary pulsars. Four pulsars exhibit DM exceeding $1000~{\rm pc~cm^{-3}}$, with J1839$-$0558t having the highest value in our sample at $\sim 1271~{\rm pc~cm^{-3}}$. These high-DM pulsars contribute valuable data points for the scattering relations in the high-DM regime. Among our discoveries, PSRs~J1836$-$0552t and J1847$-$0624t show strong nulling behavior, with nulling fractions of $\sim 86\%$ and $\sim 97\%$, respectively, placing them among RRAT-like pulsars. Their non-detection in our FFT-based periodicity search underscores the critical role of FFA and single-pulse search techniques in uncovering faint or highly intermittent pulsars. This in turn suggests that a fraction of the Galactic pulsar population may remain hidden due to the sensitivity and methodological limitations of previous surveys.

Further timing observations are needed to obtain phase-connected solutions for all new pulsars. This will reveal whether J1832$-$0901t and J1843$-$0524t are binary or isolated, and will better characterize the sporadic emission of J1836$-$0552t and J1847$-$0624t. Although we limited our search sky coverage for this study, a systematic re-analysis of the entire FAST archive using FFA and single-pulse methods is essential. This would uncover pulsars missed by FFT-based searches, yielding a more complete census of faint and intermittent pulsars.

\begin{acknowledgments}
\quad We are grateful to the referee for valuable comments that helped improve the manuscript. As this work is entirely based on publicly released FAST observations, we gratefully acknowledge all the principal investigators who obtained the original data. 
This work made use of the data from the Five-hundred-meter Aperture Spherical radio Telescope (FAST, \url{https://cstr.cn/31116.02.FAST}). FAST is a Chinese national mega-science facility, operated by National Astronomical Observatories, Chinese Academy of Sciences.
This work was supported by National Natural Science Foundation of China (NSFC) under grant Nos.~123B2045, 12041301, 12121003, 12203051 and 12273010, and by the National Key Research and Development Program of China under grant No.~2021YFA0718500.
\end{acknowledgments}

\facilities{FAST.}

\software{
\texttt{DSPSR} \citep{vanStraten+2011},
\texttt{FETCH} \citep{FETCH},
\texttt{PRESTO} \citep{Ransom+2011}, 
\texttt{PrestoZL} \citep{Mao+2025},
\texttt{PSRCHIVE} \citep{psrfits},
\texttt{Psrqpy} \citep{psrqpy},
\texttt{PulsarX} \citep{PulsarX}, 
\texttt{PyGEDM} \citep{pygedm}, 
\texttt{TransientX} \citep{TransientX} and 
\texttt{YOUR} \citep{Aggarwal+2020}.}

\bibliography{ref.bib}{}
\bibliographystyle{aasjournalv7}

\end{document}